\documentclass[aps,amssymb,amsmath,prb,reprint, showpacs]{revtex4-1}
\usepackage{times}

\usepackage[pdftex]{graphicx}
\usepackage{amsmath}
\usepackage{bm}

\usepackage{upgreek}
\usepackage[usenames]{color}
\usepackage[normalem]{ulem}

\usepackage{scalefnt}

\begin{document}


\title{Probing the electrodynamic local density of states with magnetoelectric point scatterers}

\date{January 10, 2013}

\author{Andrej Kwadrin}
\email{kwadrin@amolf.nl}
\affiliation{Center for Nanophotonics, FOM Institute AMOLF, Science Park 104, 1098 XG Amsterdam, The Netherlands}

\author{A. Femius Koenderink}
\affiliation{Center for Nanophotonics, FOM Institute AMOLF, Science Park 104, 1098 XG Amsterdam, The Netherlands}

\begin{abstract}
In a scattering experiment, the induced dipole moments of a magnetoelectric point scatterer in response to driving fields are given by its polarizability tensor $\bm{\alpha}$. Its linewidth will be dictated by the local density of optical states (LDOS) at the scatterer's position. To retrieve the magnetoelectric cross coupling components of $\alpha$ for an archetypical magnetoelectric scatterer---a split ring resonator---we study the frequency dependent extinction cross section $\sigma_\mathrm{ext}$ as a function of distance to an interface.  Rather than following a purely electric or purely magnetic LDOS, we find a dependence which reflects the interplay of both dipole moments in a 'mixed' magnetoelectric LDOS. For a strongly magnetoelectric cross-coupled microwave scatterer, we compare analytical point dipole with finite element method calculations.
\end{abstract}


\maketitle

\section{Introduction}
While electricity and magnetism are inextricably connected in optics, the interaction of light with matter is generally considered to be almost entirely mediated by the photon's electric field and the electric polarizability of matter~\cite{Landau}. In the last decade, this paradigm has shifted with the emergence of the field of metamaterials \cite{Pendry2000,Shalaev2007,Soukoulis2011}. In this field of research, complete control over the flow of light is promised by transformation optics \cite{Pendry2012}, provided one can engineer arbitrary AC (optical frequency $\omega$) permittivity $\epsilon(\omega)$ and permeability $\mu(\omega)$ of the medium it passes through. To reach this goal, many workers nanostructure materials that intrinsically have $\mu=1$ and $\epsilon \neq 1$ to create effective media that \emph{spoof} a magnetic response $\mu$. The workhorse object in the field is the so called split ring resonator \cite{Katsarakis2004,Garcia-Garcia2005,Rockstuhl2006,Rockstuhl2006a,Enkrich2005,Klein2006,Rockstuhl2007,Husnik2008,Sersic2009,Liu2009,Soukoulis2007}, a metallic ring of an overall size of around $\lambda/10$ with a single cut that provides a magnetic response through a circulating charge mode that corresponds to an LC-resonance. At telecom frequencies, experiments indicate that split rings, and similar metamaterial building blocks, have a strong magnetic polarizability of several times their physical volume \cite{Enkrich2005,Husnik2008,Sersic2009}. Therefore, it is possible to induce a strong magnetic dipole in such scatterers upon driving with the incident magnetic field of light. At the same time, metamaterial scatterers often feature a strong electric, and a so-called 'magnetoelectric' polarizability, whereby electric driving sets up a strong magnetic response and vice versa \cite{Sersic2011}.

Parallel to the development of metamaterials, interest has recently emerged in engineering magnetic fluorescent transitions \cite{Karaveli2010,Karaveli2011,Dodson2012,Taminiau2012}. Just as the response of materials to light is dominated in nature by $\epsilon$, the fluorescence of deeply subwavelength objects like atoms, quantum dots and molecules, is usually entirely dominated by electric dipole transitions \cite{NovotnyHecht}. Thus, researchers in the field of spontaneous emission control conventionally take the local density of optical states (LDOS), that quantifies how many photon states are available for an emitter to decay into, as strictly meaning the local density of \emph{electric field} vacuum fluctuations. This electric field LDOS not only governs spontaneous emission, but is also commonly associated to, e.g., light generation by cathodoluminescence \cite{GarciadeAbajo2010,Sapienza2012}, or the radiative damping of plasmonic, i.e., purely electrically polarizable, scatterers as measured by Buchler et al. \cite{Buchler2005}. However, recent work by Taminiau et al. \cite{Taminiau2012} has revealed that in some rare earth elements magnetic transition dipoles can be sufficiently strong to be cleanly observed. Those transitions sample a different LDOS, namely, the local density of \emph{magnetic field} vacuum fluctuations.

In this Paper, the two developments described above come together in a single question and its answer. If a scatterer such as a split ring is indeed a magnetic, or even a magnetoelectric scatterer of mixed electric-magnetic character, which LDOS actually sets the radiative linewidth? To formalize this question, we ask how radiation damping affects a split ring if we abstract it to a point scatterer with a formally $6\times6$ polarizability~\cite{Lindell1994,Garcia-Garcia2005} of the form
\begin{equation}
\begin{pmatrix} \bm{p} \\ \bm{m} \\ \end{pmatrix}=\begin{pmatrix} \bm{\alpha}_E & \bm{\alpha}_C \\ \bm{\alpha}_C^T & \bm{\alpha}_H \\ \end{pmatrix}\begin{pmatrix} \bm{E} \\ \bm{H} \\ \end{pmatrix}.
\label{eq:PolarizabilityTensor}
\end{equation}
Here, the electric response to electric fields and magnetic response to magnetic fields are given by the $3\times3$ tensors $\bm{\alpha}_E$ and $\bm{\alpha}_H$, respectively. The off-diagonal quantifies how strongly a magnetic dipole can be set up by an electric field and vice versa. A polarizability such as Eq.~(\ref{eq:PolarizabilityTensor}) contains very nontrivial features, such as optical activity, pseudochirality, and handed radiation patterns, depending on the amount of magnetoelectric cross-coupling $\bm{\alpha}_C$\cite{Plum2009,Sersic2011,Sersic2012}.  In how far this polarizability truly describes experiments is a matter of debate. Matching of far-field transmission spectra of periodic arrangements of such metamaterial scatterers to a lattice model for point dipoles is excellent \cite{Sersic2012Thesis} but one may wonder if the dipole picture stands up to scrutiny in a near-field experiment.   A particular near-field experiment would be to test if a split ring responds to the LDOS, a quantity specific to dipole transitions and scattering. In this paper, we first answer the question how a split ring's radiative linewidth is modified by the LDOS, and show that in addition to the electric and magnetic LDOS, a third quantity emerges in the form of a magnetoelectric LDOS. Secondly, we show that controlled variations in LDOS should allow one to measure  the magnitude and test the conceptual validity of the point scatterer's polarizability. Finally, we benchmark our proposal to finite element calculations.

\begin{figure}
	\centering
		\includegraphics[width=0.30\textwidth]{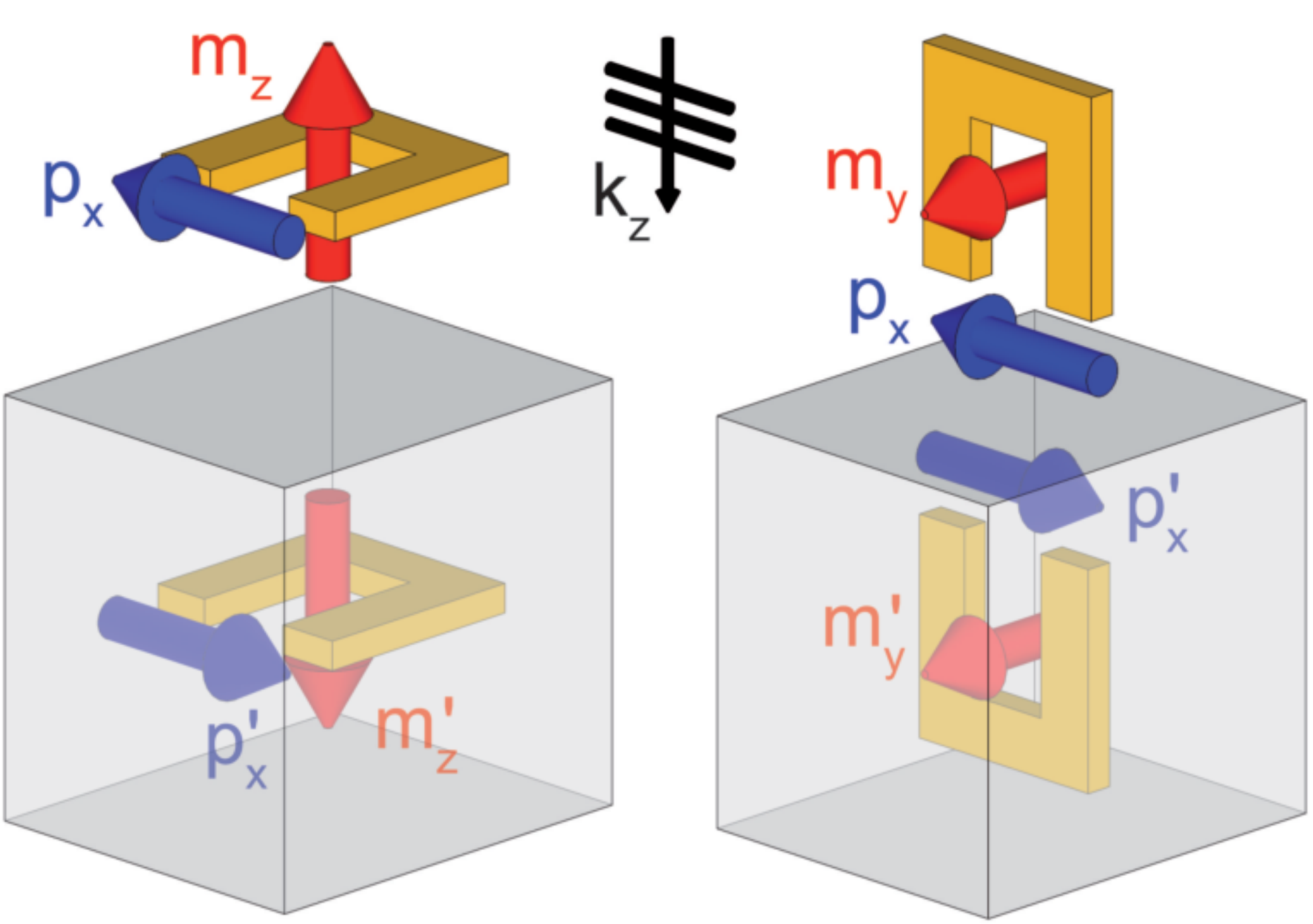}
	\caption{Split ring resonators with two distinct orientations are placed above a perfect mirror. In the point dipole picture each split ring is described by an electric dipole $\bm{p}$ along the split ring gap (blue) and a magnetic dipole $\bm{m}$ pointing out of the split ring plane (red). The mirror images for both split ring orientations are depicted together with their respective image dipoles $\bm{p'}$ and $\bm{m'}$.}
	\label{fig:Concept}
\end{figure}
\section{Magnetic LDOS near metallic and dielectric interfaces}
The effect of the LDOS on a point scatterer is well understood by first considering a polarizable dipole in front of a perfect mirror. As first demonstrated in a groundbreaking experiment by Buchler et al. \cite{Buchler2005}, the scattering resonance of a plasmon particle can be modified in width by mechanically approaching a planar reflective substrate. This effect can be interpreted  in exactly the same manner as the explanation usually given to Drexhage's experiment on the radiative transition rate of a fluorophore near a mirror~\cite{Drexhage1970,Amos1997,Snoeks1995,Leistikow2009,Kwadrin2012}.  The scattering resonance  carries an electric dipole that hybridizes with its mirror image, which for a dipole moment parallel (perpendicular) to a perfect mirror interface has reverse (identical) orientation according to image charge analysis (Fig.~\ref{fig:Concept}). The two electric dipoles together correspond to either a subradiant or superradiant configuration, depending on dipole orientation and distance. Consequently, the radiative linewidth oscillates with distance to the mirror in proportion to the electric LDOS. It is not immediately obvious that this method can be useful to also probe linewidth changes in objects with both an electric and magnetic dipole moment. While the rule for choosing the image dipole orientation reverses in the magnetic case, compared to the electric case, one should also consider that in a split ring the electric and magnetic dipole are  at 90$^\circ$ relative orientation. In most experiments, the magnetic dipole is perpendicular to the substrate (labelled $z$-oriented from hereon), while the electric dipole is in-plane ($x$-oriented). An image dipole analysis assuming a perfect mirror predicts essentially no discernible difference between the linewidth of an in-plane electric dipole and an out-of-plane magnetic dipole. That electric and magnetic LDOS have essentially the same dependence for the perpendicular dipole orientations in a mirror charge analysis was also noted by Karaveli et al.~\cite{Karaveli2010}.  Discerning magnetic and electric LDOS effects hence either implies that one rotates the scatterer (Fig.~\ref{fig:Concept}, right) to have both electric and magnetic dipoles in plane, or requires that one finds an LDOS with distinct electric and magnetic spatial dependence for the desired polarizations.

\begin{figure}
	\centering
		\includegraphics[width=0.45\textwidth]{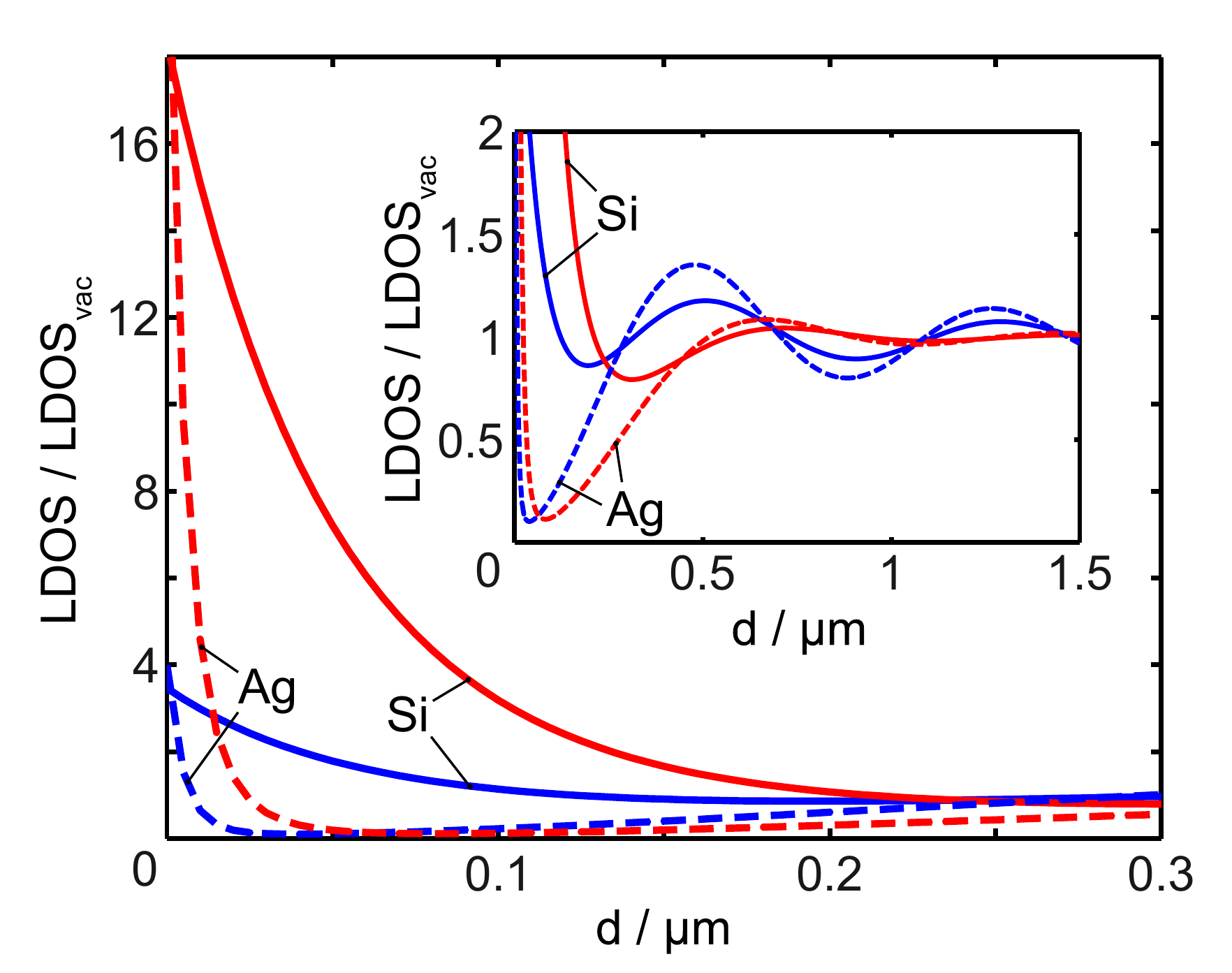}
	\caption{Relative electric ($x$-oriented dipole $\bm{p}$, blue) and magnetic ($z$-oriented dipole $\bm{m}$, red) LDOS for distance d from a vacuum/Ag (dashed lines) and vacuum/Si (solid lines) interface at a vacuum wavelength of $1.5\,\upmu$m. Especially at close distances $d<0.2\,\upmu$m, a vacuum/Si interface provides a higher LDOS contrast of $x$-oriented electric dipoles and $z$-oriented magnetic dipoles than a vacuum/Ag interface.}
	\label{fig:AgvsSiRelativeLDOS}
\end{figure}

In Fig.~\ref{fig:AgvsSiRelativeLDOS}, we plot the electric and magnetic LDOS for vacuum/Si and vacuum/Ag interfaces for both parallel and perpendicular dipole orientations with the aim of obtaining a large difference between the LDOS for $x$-oriented electric dipoles and $z$-oriented magnetic dipoles. To specify the calculation method for the LDOS at a position $\mathbf{r}$ above the interface, we calculated the imaginary part of the Green function $\bm{G}(\mathbf{r},\mathbf{r})$, as described by Novotny and Hecht \cite{NovotnyHecht}, using the complex wave vector integration contour of Paulus et al. \cite{Paulus2000}. From hereon we suppress the argument of the Green function. We generalize the calculation to encompass the electric LDOS $\mathrm{Im} \bm{G}_{EE}$, the magnetic LDOS $\mathrm{Im} \bm{G}_{HH}$ and the crossed Green dyadic $\mathrm{Im} \bm{G}_{EH}$, as specified in the appendix. We take $\epsilon=12.11$ for silicon and $\epsilon=-121.53+3.10i$ for silver, as tabulated for the resonant wavelength of $1.5\,\upmu$m \cite{Johnson1972,Aspnes1983}, typical for $200\,\mathrm{nm}\times200\,$nm split rings, and plot LDOS normalized to the LDOS in vacuum (Fig.~\ref{fig:AgvsSiRelativeLDOS}). For the vacuum/Ag interface we observe that the magnetic $z$-oriented and electric $x$-oriented LDOS are quite similar in magnitude, except within $50\,$nm of the interface. As anticipated from the perfect mirror intuition, a dielectric interface is advantageous in providing a higher contrast of magnetic electric LDOS contrast compared to a metal. Continuity conditions on $E_{\parallel}$ and $H_{\parallel}$  ensure that the electric and magnetic LDOS are highly distinct. The range over which a large distinction remains away from the interface extends well into the regime beyond the first oscillations in LDOS at $200\,$nm, as shown in Fig.~\ref{fig:AgvsSiRelativeLDOS}. Therefore, scanning the separation distance between split ring and interface allows to independently vary the electric and magnetic LDOS over a substantial range.

\section{Radiative linewidth near an interface}
Now we proceed to examine the radiative linewidth of a point scatterer described by a magnetoelectric polarizability, as proposed by Sersic et al. \cite{Sersic2011}, which is based on the static polarizability introduced by  Garcia-Garcia et al.~\cite{Garcia-Garcia2005} generalized to include radiation damping. The induced dipole moments of a scatterer in vacuum are entirely set by its full $6\times6$ dynamic polarizability tensor $\alpha^{\mathrm{dyn}}_{\mathrm{free}}$ that is of the form Eq. (\ref{eq:PolarizabilityTensor}). In a quasistatic description of the scatterer, one starts from an LC circuit to obtain a static polarizability $\alpha^{\mathrm{stat}}_{\mathrm{free}}$ that consists of a  Lorentzian frequency dependence ${\cal L(\omega)}=\frac{\omega_0^2
V}{\omega_0^2-\omega^2-i\omega\gamma}.$  (resonant at the $LC$ frequency $\omega_0$ , damping rate $\gamma$ set by the Ohmic resistance $R$) multiplying a frequency independent  $2\times2$ matrix
\begin{equation}
\bm{\alpha}^{\mathrm{stat}}_{\mathrm{free}}={\cal L(\omega)} \begin{pmatrix} 
\eta_{E,xx} &  i\eta_{C,xz} \\  -i \eta_{C,zx} &  \eta_{H,zz} \end{pmatrix}
\end{equation}
where $\eta_E, \eta_H, \eta_C$ are real parameters simply set by geometry.  
For an ideal, infinitely thin SRR all other elements of the $6\times6$ polarizability are zero. To end up with an energy conserving scatterer, a radiation damping term must be added:
\begin{equation}
{\bm{\alpha}^{\mathrm{dyn}}_{\mathrm{free}}}^{-1}={\bm{\alpha}^{\mathrm{stat}}_{\mathrm{free}}}^{-1}-i \mathrm{Im} \bm{G}.
\label{eq:DynamicAlpha}
\end{equation}
where $\bm{G}$ is the 6$\times$6 Green function.
In vacuum the correction amounts to the usual radiation damping term $i \mathrm{Im} \bm{G}=2/3ik^3 \mathbb{I}$, where $k=\omega /c$, and $\mathbb{I}$ the identity matrix, that is standard in the field of plasmonics.
In front of the interface, however, the Green function is modified according to Fig.~\ref{fig:AgvsSiRelativeLDOS}.  For the effectively 2$\times$2 polarizability of the split ring, the relevant $\mathrm{Im}\mathbf{G}$ tensor is also only $2\times 2$, containing on the diagonal only the electric LDOS for $x$-oriented dipoles $\mathrm{Im} \bm{G}_{EE,xx}$, and   the magnetic LDOS for $z$-oriented dipoles  $\mathrm{Im} \bm{G}_{HH,zz}$, while the off diagonal contains  $\mathrm{Im} \bm{G}_{EH,xz}$.

\begin{figure*}
	\centering
		\includegraphics[width=0.95\textwidth]{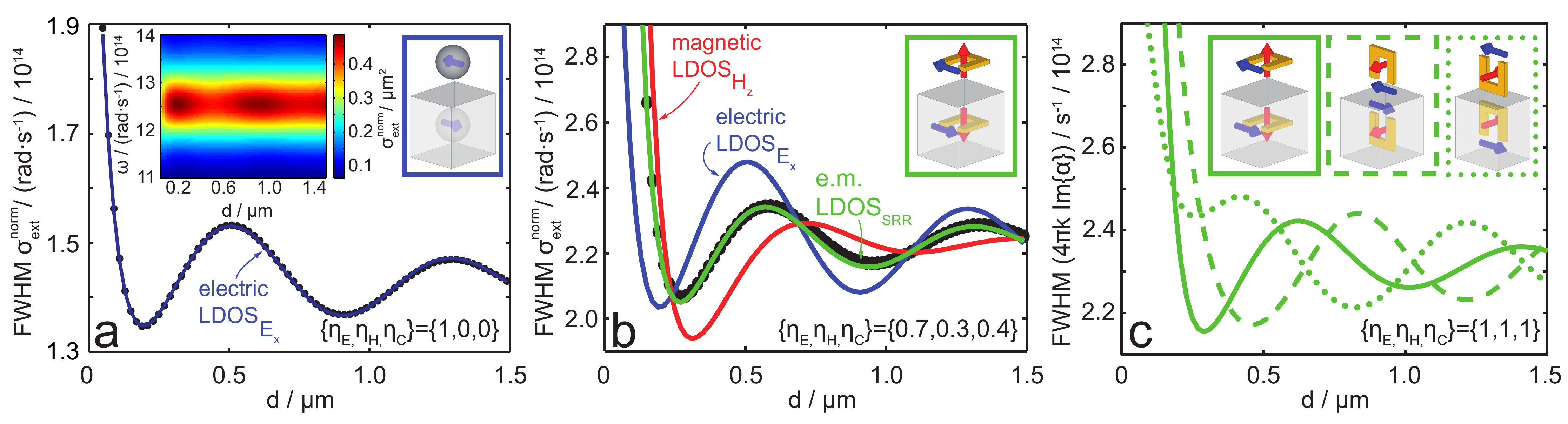} 
	\caption{(a) Analytical calculation of the FWHM-linewidth of the extinction cross section normalized to driving field intensity as a function of distance $d$ to a vacuum/Si interface for a purely electric ($\eta_E=1,\eta_H=\eta_C=0$)  sphere resonant at 1.5~$\upmu$m wavelength, with Ohmic damping rate $\gamma=8.3\cdot10^{13}~$s$^{-1}$ and a particle volume of $V=100~$nm$^{3}$ (dots) in comparison to the purely electric LDOS lineshape for this interface. (b) The dots represent the same quantitity as before, but for realistic split ring resonator ($\eta_H=0.7,\eta_E=0.3,\eta_C=0.4$)\cite{Sersic2009} that is oriented with the SRR plane parallel to the surface. Electric, magnetic and magnetoelectric LDOS are shown as lines. (c) Maximally cross-coupled split ring resonator ($\eta_E=\eta_H=\eta_C=1$) parallel to and with the gap pointing away and pointing towards the interface, respectively.}
	\label{fig:Fig3BenchmarkAndAnalytical}
\end{figure*}

We obtain the radiative linewidth as one would measure it in an extinction experiment, by calculating the extinction cross section in the following manner. The scatterer is driven by a plane wave impinging from above, plus its Fresnel reflection due to the interface. We take the incident beam as normal to the interface with the electric field  polarized along the gap. We calculate the work done per unit cycle on the scatterer via
\begin{equation}
W= \left\langle \mathrm{Re}\bm{E}\cdot \mathrm{Re}\frac{d\bm{p}}{dt} + \mathrm{Re}\bm{H}\cdot \mathrm{Re}\frac{d\bm{m}}{dt}  \right\rangle. 
\end{equation}
Plots of the work show an oscillatory dependence with distance of the split ring to the interface, due to two effects. Firstly, the driving field forms a standing wave. Secondly, the polarizability varies with the oscillating LDOS. To obtain a true extinction cross section we  divide out the local field strength of the driving 
\begin{equation}
\sigma_\mathrm{ext}=\frac{2Z}{{\left|\bm{E}\right|}^2}W,
\end{equation}
with $Z$ the impedance of the host medium, in this case vacuum. As Fig.~\ref{fig:Fig3BenchmarkAndAnalytical}(a, inset)  shows, the retrieved extinction cross section is of Lorentzian spectral shape, and varies in width and central frequency as the scatterer is approached to the interface. We extract the resonance width, which is the sum of the radiative and absorptive damping rate of the scatterer.  

As a benchmark, Fig.~\ref{fig:Fig3BenchmarkAndAnalytical} shows the linewidth of the extinction cross section of a purely electric scattering sphere, i.e., taking $\eta_E=1$, $\eta_H=0$ and $\eta_C=0$, resonant at 1.5~$\upmu$m wavelength, with Ohmic damping rate $\gamma=8.3\cdot10^{13}~$s$^{-1}$ and a particle volume of $V=100~$nm$^{3}$ . As the scatterer approaches the interface, its extinction linewidth oscillates, and almost doubles when the scatterer is close to the interface. We find that the linewidth $\Gamma$ follows to great accuracy the dependence  $\Gamma=\Gamma_{\mathrm{abs}} + \Gamma_{\mathrm{rad}} \times \mathrm{LDOS}_{EE,xx}$. Analogous to the calibration of quantum efficiencies of fluorophores \cite{Snoeks1995,Leistikow2009,Kwadrin2012}, this dependence allows to extract the radiative and Ohmic damping rate of the particle, and thereby also the LDOS dependent albedo of the scatterer. For the sphere  studied here, the albedo in absence of the interface is $a=0.41$. This benchmark calculation shows that our model essentially reproduces the experimental observation by Buchler et al.\cite{Buchler2005}. Similarly, a calculation for a purely magnetic scatterer verifies that the damping rate of a magnetic scatterer traces the magnetic LDOS [calculation not shown].

In Fig.~\ref{fig:Fig3BenchmarkAndAnalytical}, as a measure for $\Gamma$, we examine the extinction linewidth for objects that have both an electric and a magnetic character. For demonstration purposes, we take the electric and magnetic polarizability equally large at $\eta_E=\eta_H=1$. If no cross-coupling, i.e., no bianisotropy is present in the object ($\eta_C=0$), the extinction linewidth simply traces the electric LDOS, provided excitation is normal to the sample so that the magnetic dipole is not directly driven at all [curve not shown]. As cross-coupling is introduced, and increased to its maximum value,  the extinction linewidth shifts away from the purely electric LDOS, and towards the magnetic LDOS curves. At maximum cross coupling ($\eta_C=1$), the extinction linewidth exactly traces the mean of electric and magnetic LDOS consistent with the fact that electric and magnetic dipole are equal in size.
Generally, for this geometry and excitation, the averaging is weighted by the magnitudes of the dipole moments, i.e., $|\bm{p}|^2$ and $|\bm{m}|^2$. We conclude that an experiment such as performed by Buchler et al.\cite{Buchler2005}, but applied to a split ring can indeed provide a quantitative test of the magnetic and bianisotropic dipole response of a single object, and a calibration of the magnitude of polarizability tensor elements.

\begin{figure}
	\centering
	\includegraphics[width=0.45\textwidth]{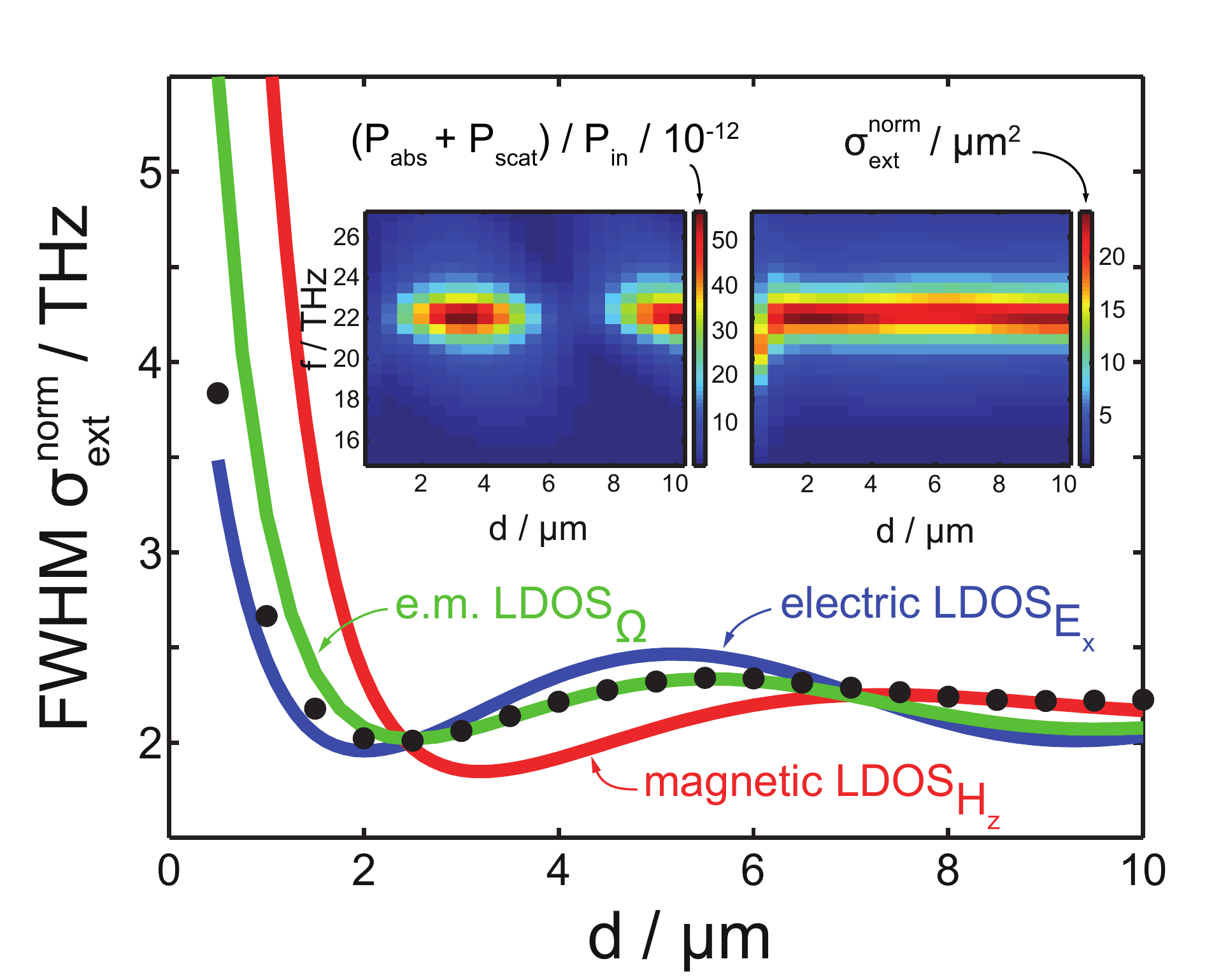}
	\caption{FWHM-linewidth of the extinction cross section normalized to the local driving field intensity for a zero arm-length omega particle oriented parallel to a Si interface as a function of distance d. Retrieved FWHM-linewidths from FEM-simulation (dots) vs. point dipole calculation (lines). Insets show extinct power (left) as well as normalized extinction cross section (right) as a function of frequency $f$ and distance $d$ as acquired from FEM-simulation.}
	\label{fig:Figure4FEMpd}
\end{figure}

\section{Magnetoelectric LDOS}
That a weighted average of electric and magnetic LDOS is obtained for a scatterer with both electric and magnetic dipole moment, may seem a likely general description of the physics describing radiative linewidths. However, we note that in a split ring the electric and magnetic dipole are \textit{coherently} coupled through the magnetoelectric coupling, and hence may also probe the off-diagonal term in $\mathrm{Im}\mathbf{G}$. Moreover, in a split ring the electric and magnetic response have a fixed phase relation, necessarily being a quarter cycle out of phase since the current inducing $\bm{m}$ and the charge separation inducing $\bm{p}$ are related by charge conservation. This coherence for instance results in a strongly handed response under certain viewing angles~\cite{Sersic2011,Sersic2012}. We predict that the coherence also affects the strength of interaction between the split ring and its mirror image in the substrate, i.e., the radiative linewidth.  In the example we examined in Fig.~\ref{fig:Fig3BenchmarkAndAnalytical}, this effect was fortuitously obscured due to the fact that the cross term $\bm{G}_{EH,xz}$ by symmetry happens to be exactly zero. Microscopically,  the radiation emitted by an $x$-oriented electric dipole in front of a planar surface does not cause any magnetic field along $z$ to be reflected to the split ring to provide back action.  If we rotate the split ring to stand with arms up, or point with arms down, cross coupling becomes important as $\bm{G}_{EH,xy} \neq 0$.   Microscopically, this indicates that an $x$-oriented electric dipole will receive a $y$-oriented magnetic field as parts of its reflection in the interface, which in turn will drive the object if it has a $y$-oriented magnetic polarizability. In Fig.~\ref{fig:Fig3BenchmarkAndAnalytical} we report the radiative linewidth for a point scatterer oriented to exactly represent this case, i.e., that of a split ring that stands up with its legs normal to the plane.  The radiative linewidth in this case does not trace a weighted combination of magnetic and electric LDOS. It will depend on $\bm{G}_{EH,xy}$ instead.  Remarkably, this dependence is different for the split ring pointing upwards to the split ring pointing downwards although the object has the  \emph{same} electric and magnetic polarizability.  The only difference is the sign of the cross coupling polarizability, i.e., whether the quarter cycle phase difference between magnetic and electric response is a  lag or an advance.  Thus   the fact that radiation reaction is a coherent effect means that the linewidth provides a direct way to measure phase relations between polarizabilities, and not only absolute values. For instance, one could measure if the quarter wave phase difference between $\bm{\alpha}_E, \bm{\alpha}_H$ on the one hand and $\bm{\alpha}_C$ on the other hand, that is generally surmised from quasistatic ciruit theory for split rings, in fact carries over to real scatterers that are not negligibly small compared to the wavelength and that are not composed of ideal conductors.   To our knowledge, this is the first proposition that a new  property of the structure that can potentially be engineered independently of the well-known electric LDOS and the recently evidenced magnetic LDOS, may enter radiative linewidth modifications for dipole objects. We term this a magnetoelectric LDOS effect.
 
\section{Finite element modelling example}
Our predictions for the effect of the magnetic and magnetoelectric LDOS on the linewidth of metamaterial scatterers are all subject to the assumption that the scatterers can actually be described in a point dipole picture. An important question that is yet to be tested in experiment and simulations is if this assumption holds at all for metamaterial scatterers, and if so for what classes of optical experiments. Therefore, we perform a numerical experiment and compare the radiative linewidth found from finite element method calculation with the point dipole predictions. To optimally discriminate for magnetic LDOS effects, we choose an omega-shaped particle that earlier surface integral equation simulations(SIE\cite{Sersic2012,Kern2009}) predict to have a very large magnetic and magnetoelectric response. In Ref.~\cite{Sersic2012}, the polarizability was quantitatively retrieved by projecting calculated scattered fields for the scatterer held in free space on vector spherical harmonics. We have performed full-field finite element calculations of the omega-shaped scatterer, which is resonant in the microwave regime, again above a  high-index substrate ($n=3.5$). The scatterer geometry is that of a 60 nm thin flat loop of inner radius 0.74~$\upmu$m  and outer radius $1.19~\upmu$m radius. Before the loop closes, the arms smoothly curve to be parallel over a length of 390~nm, leaving  a gap of 520~nm across. As material we use a Drude model for gold ($\epsilon(\omega)=\epsilon_b-\omega_p^2/(\omega(\omega+i\gamma))$ with $\epsilon_b=9.54, \omega_p=2.148\cdot10^{15}~$s$^{-1}$ and $\gamma=0.0092\omega_p$. We employ the commercial COMSOL 3D FEM solver with elements of quadratic order and a grid finesse down to 5~nm. Perfectly matched layers enclose a cylindrical simulation domain (cylinder axis normal to the substrate) that extends 10~$\upmu$m around the scatterer. We perform a total field-scattered field simulation where, as in the point dipole model, the scatterer is excited by the superposition of a plane wave and its Fresnel reflection. We extract extinct power as the sum of scattered power (obtained from a near-field flux integral over a surface enclosing the particle) and  absorbed power, and normalize extinct power to the local driving strength.  This procedure was tested on Mie scatterers to give cross sections to better than 1\%.

Extinction spectra show a Lorentzian linewidth at all separations, with a varying width and a slightly varying center frequency around a wavelength of 13.5$~\upmu$m. The center frequency varies because Eq.~\ref{eq:DynamicAlpha} in its most general form also contains $\mathrm{Re} \bm{G}$, corresponding to a real frequency shift due to the hybridization of the scatterer with its mirror image. We focus on the linewidth, plotted in Fig.~\ref{fig:Figure4FEMpd}. Evidently, the linewidth shows oscillations increasing in amplitude when approaching the interface, and is twofold larger close to the interface than away from it. To move beyond this qualitative resemblance with the point dipole prediction, we also plot the linewidth found from point dipole calculations. No adjustable parameters are used for the comparison, as we insert the polarizability values extracted from SIE calculations in Ref.~\onlinecite{Sersic2012}, which are characterized through $\alpha_H/\alpha_E=0.3511$ and $\alpha_C/\alpha_E=0.596$. We note that the object has an on-resonance electric polarizability $|\alpha_E|=3.9~\upmu$m$^3$, approximately 30 times the particle volume.
The point dipole model is seen to satisfactorily agree with the linewidth in simulations. Thereby, we establish that the point dipole approach not only describes far-field measurements on  arrays of split rings, but  is also directly applicable for split rings in the near-field of structures that modify LDOS. We propose that the residual deviations contain interesting physics to pursue. Firstly,  whether a metamaterial scatterer actually traces the magnetic and magnetoelectric LDOS can be seen as a fundamental test in the discussion in  how far a spoof magnetic scatterer is actually  a true magnetic scatterer. Secondly, if one accepts that a scatterer that largely radiates according to the magnetic and magnetoelectric LDOS is a bona fide magnetic dipole, one can assess on basis of the residual deviations between simulation and analytical model in how far multipolar corrections are important.

\section{Conclusion}
To conclude we have examined the dependence of the radiative linewith of split ring scatterers on their distance to an interface that modifies the electric LDOS, the magnetic LDOS and a new quantity that we term magnetoelectric LDOS. We propose that this linewidth, i.e., the backaction of the field radiated by the scatterer on itself, can serve as a calibration probe of the complex polarizability tensor and as a fundamental test of the proposed dipolar nature of metamaterial scatterers.  Of particular note is the concept of magnetoelectric LDOS, whereby the electric dipole of an object radiates magnetic field that back-acts on the magnetic dipole. It is an interesting question to explore whether such a  magnetoelectric LDOS will also affect  quantum mechanical transitions.  While in the recent breakthroughs by Taminiau et al. and Karaveli et al. magnetic-only transitions in rare earth ions are enhanced~\cite{Karaveli2010,Karaveli2011,Taminiau2012}, it is an open question if transitions with a clear simultaneous electric and magnetic character can be found  Conversely, we propose that coupling single emitters with a purely electric response to magnetoelectric scatterers may allow to spoof quantum mechanical transitions with a magnetoelectric character, as antennas tend to imbue their polarization characteristics on emitters. Such emitters woud likely have interesting chiral properties, since magnetoelectric cross coupling implies optical activity~\cite{Barron2004,Tang2010}.
 
\begin{acknowledgments}
We thank Felipe Bernal Arango for sharing surface integral equation calculations, and Anouk de Hoogh for sharing COMSOL experience. This work is part of the research programme of the Foundation for Fundamental Research on Matter (FOM), which is part of The Netherlands Organisation for Scientific Research (NWO). A.F.K. acknowledges a NWO-Vidi fellowship.
\end{acknowledgments}

\appendix*
\section{Green function}

To calculate LDOS, we require the $6\times 6$ dyadic Green function near a planar interface where source $\mathbf{r}'$ and observation point $\mathbf{r}$ are in the same halfspace.We separate the Green function in a free part (in absence of an interface) and a reflected part
$$\mathbf{G}(\mathbf{r},\mathbf{r'})=\mathbf{G}_\mathrm{free}(\mathbf{r},\mathbf{r'}) + \mathbf{G}_\mathrm{reflected}(\mathbf{r},\mathbf{r'})$$
with 
\begin{equation}
\mathbf{G}_\mathrm{free}(\mathbf{r},\mathbf{r}') =
\begin{pmatrix}
\mathbb{I} k^2 +\nabla\nabla &
-ik\nabla \times  \\
 ik\nabla \times   &  \mathbb{I} k^2
+\nabla\nabla \\
\end{pmatrix} G(\mathbf{r},\mathbf{r}')
\label{eq:derive}
\end{equation}
where $k=\omega n/c$ is the wave number in the medium of index $n$ that contains both $\mathbf{r}$ and $\mathbf{r'}$, $c$ is the speed of light, and $G(\mathbf{r},\mathbf{r}')$ is the scalar Green funtion. The reflected part of the green function reads 
\begin{eqnarray}
\mathbf{G}_\mathrm{reflected}(\mathbf{r},\mathbf{r}') &=&
\frac{i k^2}{2}\int_0^\infty k_{||}
d{k}_{||} [j_0(k_{||}R) \mathbf{M}_0 +   \nonumber \\ & & \quad j_1(k_{||}R) \mathbf{M}_1   +j_2(k_{||}R) \mathbf{M}_2] e^{ik_z Z}   \nonumber \\
\end{eqnarray}
where if $\mathbf{r}=(x,y,z)$ and $\mathbf{r'}=(x',y',z')$ we define cylindrical coordinates through 
$(R\cos\phi, R\sin\phi, Z)=(x-x',y-y',z+z')$.  With $k_z$ we denote $\sqrt{k^2-||k_{||}||^2}$, while $j_n(x)$ is the Bessel function of order $n$.  The $6\times6$ matrices $\mathbf{M}_i$  contain the $k_{||}$ dependent Fresnel reflection coefficients $r_s$ and $r_p$  for $s$ and $p$ polarization.  In detail:
\begin{widetext}
$$\mathbf{M}_0=
\begin{pmatrix}
r_s/k_z  - r_p k_z 	& 0 & 0 		& 0  & {r_p - r_s}&  0 \\
0 		& r_s/k_z -r_p k_z & 0              & {r_s -r_p} & 0 & 0 \\
0  	& 0 &{ 2k_{||}^2 r_p/k_z}  			&  0 & 0 & 0 \\
0 & r_p-r_s  & 0  				&  r_p/k_z-r_s k_z& 0 & 0 \\
r_s - r_p &  0 & 0                                      &  0 & r_p/k_z-r_s k_z & 0 \\
 0  & 0  & 0                                          & 0  &   0         & {2 r_sk_{||}^2/k_z}\\
\end{pmatrix}
$$\end{widetext}
and
\begin{widetext}
$$
\mathbf{M}_1=2ik_{||}
\begin{pmatrix}
0 & 0 &  {-  r_p    \cos\phi    } & 0 & 0 & {-r_s/k_z    \sin\phi  }\\
0 & 0 &{ - r_p      \sin\phi }    & 0 & 0 &  {r_s/k_z   \cos \phi} \\
 r_p    \cos\phi &    r_p    \sin\phi & 0    &   {r_p/k_z    \sin\phi} &      { -r_p /k_z\cos\phi} & 0 \\
0 & 0 & 	{  r_p/k_z   \sin\phi }    & 0 & 0 & {- r_s   \cos \phi }\\
0 & 0 & { - r_p/k_z   \cos\phi    } & 0 & 0 & {- r_s   \sin \phi }\\
-  r_s/k_z   \sin\phi &   r_s/k_z   \cos\phi & 0 &    r_s   \cos\phi &    r_s    \sin\phi &  0 \\
\end{pmatrix}
$$
\end{widetext}
and finally also
\begin{widetext}
$$\mathbf{M}_2= 
\begin{pmatrix}
(r_s/k_z    + r_pk_z)\cos2\phi  &( r_s/k_z +r_pk_z)\sin2\phi  & 0     &  (r_s + r_p)\sin 2\phi  &  {-}  ( r_s    + r_p)\cos2\phi & 0 \\
(rs/k_z+ r_pk_z)\sin2\phi     & -(r_s/k_z+r_pk_z) \cos2\phi & 0   &  {-}(r_s+r_p)\cos2\phi &    {-}( r_s +r_p) \sin2\phi  & 0 \\
0                          &     0                        & 0   &    0                &           0             &  0 \\
-(r_s+r_p)\sin2\phi        & ( r_s+r_p)\cos2\phi         & 0   & (r_sk_z+r_p/k_z) \cos2\phi   & (r_sk_z+r_p/k_z)\sin2\phi & 0 \\
(r_s+r_p)\cos2\phi	    & (r_s+r_p) \sin2\phi           &  0  & (r_sk_z+r_p/k_z)\sin2\phi       & -(r_sk_z+r_p/k_z)\cos2\phi & 0 \\
 0                         &       0                    &   0  &  0 		& 0       &0   \\
\end{pmatrix}
  $$.
\end{widetext} 
Throughout we have used the units of Ref.~\cite{Sersic2011}

\bibliography{AKwadrin_AFKoenderink_LDOSprobe}
\end{document}